%% file: backdoorBerge.tex
\documentclass[letterpaper,11pt]{article}

\newcommand{\longversion}[1]{#1}
\newcommand{\shortversion}[1]{}
\newcommand{\longshort}[2]{\longversion{#1}\shortversion{#2}}

\shortversion{\documentclass[runningheads]{llncs}}

\longversion{\usepackage{amsthm}}

\usepackage{amsmath,amssymb,amstext}

\usepackage{url} 
\usepackage{booktabs}
\usepackage{threeparttable}
\usepackage{tabularx}

\usepackage{tikz}
\usepackage{boxedminipage}
\usepackage{xspace}
\usepackage{wasysym}

\shortversion{
 \usepackage[footnotesize,skip=0pt,belowskip=-10pt]{caption}
}
\usepackage{subfig}

\longversion{
 \usepackage{vmargin}
 \setpapersize{USletter}
 \setmarginsrb{1in}{1in}{1in}{0.5in}{0cm}{0cm}{0.25in}{0.25in}
}

\longshort{
 
 \newtheorem{lemma}{Lemma} 
 
 \newtheorem{theorem}{Theorem}

 \newtheorem{observation}{Observation}
 
  \newcommand{\myqed}{}
}{
 \spnewtheorem{observation}{Observation}{\bfseries}{\itshape}
 \spnewtheorem{myrule}{Rule}{\bfseries}{\itshape}
  \newcommand{\myqed}{\qed}
}

\usepackage{thm-restate} 

\newcommand{\hy}{\hbox{-}\nobreak\hskip0pt}

\newcommand{\CCard}[1]{\|#1\|}

\newcommand{\cC}{\mathcal{C}} \let\CCC\cC

\newcommand{\cF}{\mathcal{F}}

\newcommand{\cS}{\mathcal{S}}

\newcommand{\NP}{\text{\normalfont NP}}

\newcommand{\FPT}{\text{\normalfont FPT}}
\newcommand{\XP}{\text{\normalfont XP}}
\newcommand{\W}[1][xxxx]{{\normalfont \text{W}[#1]}}

\newcommand{\fpt}{fixed-pa\-ra\-me\-ter trac\-ta\-ble\xspace}

\newcommand{\set}[1]{\left\{ #1 \right\}}
\newcommand{\myiff}{iff\xspace}
\newcommand{\myWlog}{W.l.o.g.\xspace}

\newcommand{\ie}{i.e.\xspace}
\newcommand{\etal}{\emph{et al.}\xspace}

\newcommand{\var}{{\normalfont \textsf{var}}}
\newcommand{\cla}{{\normalfont \textsf{cla}}}
\newcommand{\lit}{{\normalfont \textsf{lit}}}
\newcommand{\truef}{{\normalfont \textsf{true}}} 
\newcommand{\true}{1} 
\newcommand{\false}{0} 

\newcommand{\inc}{{\normalfont \textsf{inc}}}
\newcommand{\slit}{{\normalfont \textsf{slit}}}
\newcommand{\tw}{{\normalfont \textsf{tw}}}
\newcommand{\sign}{{\normalfont \textsf{sign}}}
\newcommand{\signp}{\ensuremath{+}}
\newcommand{\signm}{\ensuremath{-}}

\newcommand{\class}[1]{\text{\text{\normalfont\sc  #1}}}
\newcommand{\HORN}{\class{Horn}}
\newcommand{\RHORN}{\class{RHorn}}
\newcommand{\TWOCNF}{\class{2\hy CNF}}
\newcommand{\CLU}{\class{Clu}}
\newcommand{\FOREST}{\class{Forest}}
\newcommand{\UP}{\class{UP}}

\newcommand{\BDS}{BDS\xspace}
\newcommand{\BDSs}{BDSs\xspace}

\newcommand{\BDSsfirst}{Back\-door Sets (\BDSs{})\xspace}
\newcommand{\BT}{Back\-door Tree\xspace}

\newcommand{\acyclic}{a\-cy\-clic\xspace} 
\newcommand{\Forest}{\textsc{Fo\-rest}\xspace}
\newcommand{\FBDS}{\Forest{}\hy \BDS}
\newcommand{\FBDSs}{\Forest{}\hy \BDSs}
\newcommand{\wBDS}{weak \FBDS}
\newcommand{\wBDSs}{weak \FBDSs}
\newcommand{\WBDS}{Weak \FBDS}
\newcommand{\WBDSs}{Weak \FBDSs}
\newcommand{\WBDSpb}{\textsc{\WBDS De\-tec\-tion}\xspace}
\newcommand{\sBDS}{strong \FBDS}
\newcommand{\sBDSs}{strong \FBDSs}
\newcommand{\SBDS}{Strong \FBDS}
\newcommand{\SBDSs}{Strong \FBDSs}
\newcommand{\FBT}{\Forest{}\hy \BT}

\newcommand{\BTpb}{\textsc{\FBT De\-tec\-tion}\xspace}
\newcommand{\SBDSpb}{\textsc{\SBDS De\-tec\-tion}\xspace}
\newcommand{\DBDSpb}{\textsc{Deletion \FBDS De\-tec\-tion}\xspace}

\newcommand{\vdcs}{ver\-tex-dis\-joint cy\-cles\xspace}
\newcommand{\fvs}{feed\-back ver\-tex set\xspace}
\newcommand{\fvss}{feed\-back ver\-tex sets\xspace}

\newcommand{\kcur}{k}
\newcommand{\kcycles}{\mathsf{cycles}(\kcur)}
\newcommand{\fcycles}{\mathsf{cycles}}
\newcommand{\kextcycles}{\mathsf{ext\hy cycles}(\kcur)}
\newcommand{\kmulti}{\mathsf{multi}(\kcur)}
\newcommand{\ksupp}{\mathsf{supp}(\kcur)}
\newcommand{\koverlap}{\mathsf{overlap}(\kcur)}
\newcommand{\kfvs}{\mathsf{fvs}(\kcur)}

\allowdisplaybreaks

\newcommand{\pbDefP}[4]{%
\noindent
\begin{center}
\begin{boxedminipage}{0.98 \columnwidth}
#1\\[5pt]
\begin{tabular}{l p{0.82 \columnwidth}}
Input: & #2\\
Parameter: & #3\\
Question: & #4
\end{tabular}
\end{boxedminipage}
\end{center}
}

\title{Backdoors to Acyclic SAT
\thanks{Research supported by the European Research Council (ERC), project COMPLEX REASON 239962.%
 \shortversion{ Proofs of statements marked with $(\star)$ are in the appendix.
  The reader who prefers reading the full version of the paper is referred to the arXiv \cite{GaspersSzeider11a}.}}}
\author{%
Serge Gaspers \and
Stefan Szeider}

\longshort{
 \date{%
 Institute of Information Systems\\ Vienna University of Technology\\ Vienna, Austria.\\
 \texttt{gaspers@kr.tuwien.ac.at}\\ \texttt{stefan@szeider.net}
 }
}{
 \institute{Institute of Information Systems, Vienna University of Technology, Vienna, Austria.\\
 \texttt{gaspers@kr.tuwien.ac.at}, \texttt{stefan@szeider.net}}
}

\begin{document}

\maketitle

\shortversion{
 \setlength{\abovedisplayskip}{3pt}
 \setlength{\belowdisplayskip}{3pt}
 \setlength{\jot}{2pt}
}

\begin{abstract}
  Backdoor sets, a notion introduced by Williams \etal in 2003, are certain
  sets of key variables of a CNF formula $F$ that make it easy to
  solve the formula; by assigning truth values to the variables in a
  backdoor set, the formula gets reduced to one or several
  polynomial-time solvable formulas. More specifically, a \emph{weak
    backdoor set} of $F$ is a set $X$ of variables such that there exits
  a truth assignment $\tau$ to $X$ that reduces $F$ to a satisfiable
  formula $F[\tau]$ that belongs to a polynomial-time decidable base
  class $\CCC$.  A \emph{strong backdoor set} is a set $X$ of variables
  such that for all assignments $\tau$ to $X$, the reduced formula $F[\tau]$
  belongs to $\CCC$.

  We study the problem of finding backdoor sets of size at most $k$ with
  respect to the base class of CNF formulas with acyclic incidence
  graphs, taking $k$ as the parameter. We show that
  \begin{enumerate}
  \item the
    detection of weak backdoor sets is $\W[2]$\hy hard in general but
    fixed-parameter tractable for $r$\hy CNF formulas, for any fixed $r\ge 3$, and 

  \item the detection of
    strong backdoor sets is fixed-parameter approximable. 
  \end{enumerate}
  Result~1 is the the first positive one for a base class that does not
  have a characterization with obstructions of bounded size.  Result~2
  is the first positive one for a base class for which strong backdoor
  sets are more powerful than deletion backdoor sets.
  
  Not only SAT, but also \#SAT can be solved in polynomial time for CNF
  formulas with acyclic incidence graphs. Hence Result~2 establishes a
  new structural parameter that makes 
  \#SAT fixed-parameter tractable and that is incomparable with known parameters such as treewidth
  and cliquewidth.

  We obtain the algorithms by a combination of an
  algorithmic version of the Erd\H{o}s-P\'{o}sa Theorem, Courcelle's model checking
  for monadic second order logic, and new combinatorial results on how disjoint cycles can
  interact with the backdoor set.
\longversion{%
  These new combinatorial arguments come into play when
  the incidence graph of $F$ has many vertex-disjoint cycles. As only few of these cycles
  can vanish by assigning a value to a variable from the cycle, many cycles need to vanish
  by assigning values to variables that are in clauses of these cycles. These external variables
  are either so rare or structured that our combinatorial
  arguments can identify a small set of variables such that any backdoor set of size
  at most $k$ contains at least one of these variables, or they are so abundant and unstructured
  that they themselves create cycles in the incidence graph in such a way that $F$ cannot have a backdoor
  set of size at most $k$.
}
  
\longversion{
}
%
 





\end{abstract}

%
%
%
%
%
%
%

\section{Introduction}

Since the advent of computational complexity in the 1970s it quickly
became apparent that a large number of important problems are
intractable~\cite{GareyJohnson79}. This predicament motivated
significant efforts to identify tractable
\longshort{subproblems within intractable problems}{special cases}.
For the propositional satisfiability problem
(SAT), dozens of such ``islands of trac\-ta\-bi\-li\-ty'' have been identified
\cite{FrancoMartin09}. Whereas it may seem unlikely that a real-world
instance belongs to a known island of tractability, it may be ``close''
to one.  In this paper we study the question of whether we can exploit
the proximity of a SAT instance to the island of
acyclic formulas algorithmically.
  
For SAT, the distance to an island of tractability (or \emph{base class})
$\CCC$ is most naturally measured in terms of the number of variables
that need to be instantiated to put the formula into~$\CCC$. Williams
\etal \cite{WilliamsGomesSelman03} introduced the term ``\emph{backdoor set}''
 for sets of such variables, and
distinguished between weak and strong backdoor sets.
A set $B$ of variables is a \emph{weak $\CCC$\hy backdoor set} of a CNF formula $F$ if
for at least one partial truth assignment $\tau:B\rightarrow \{0,1\}$, the restriction
$F[\tau]$ is satisfiable and belongs to\longversion{ the base class} $\CCC$.
\longversion{($F[\tau]$ is obtained from $F$ by removing all clauses
that contain a literal that is true under $\tau$ and by removing from
the remaining clauses all literals that are false under $\tau$.) }%
The set $B$ is a \emph{strong $\CCC$\hy backdoor set} of $F$
if for every partial truth assignment $\tau:B\rightarrow
\{0,1\}$ the restriction $F[\tau]$ belongs to $\CCC$.
\longversion{%
 The base classes considered in the
 sequel are defined in Table~\ref{tab:classes}.

 \begin{table}[tb]
 \begin{threeparttable}
 \centering
 \begin{tabularx}{\textwidth}{@{}lX@{}}
  \toprule 
  \emph{Base Class}      & \longshort{\emph{Description}}{\emph{Restriction on CNF formulas in the base class}} \\
  \midrule
  $\HORN$  & \longshort{Horn formulas, i.e., CNF formulas where each}{Each}
  clause contains at most one positive literal. \\

  $\RHORN$ & \longshort{Renamable Horn formulas, i.e.,  CNF
  formulas that}{The formula} can be made Horn by flipping literals.\\

  $\TWOCNF$ & \longshort{Krom formulas, i.e., CNF formulas where each}{Each} clause
  contains at most two literals. \\ 

  $\CLU$ & \longversion{Cluster formulas, i.e.,}  CNF
  formulas that are variable disjoint unions of hitting formulas.  A
  formula is \emph{hitting} if any two of its clauses clash in at least one
  variable.\\

  $\UP$ & CNF formulas from which the empty formula or an empty clause can be derived by
  unit propagation.\\

  $\FOREST$ &  Acyclic formulas, i.e.,  CNF
  formulas whose incidence graphs are forests. The incidence graph is the
  undirected bipartite graph on clauses and variables where a variable
  is incident with all the clauses in which it occurs.\\

  \bottomrule
 \end{tabularx}
  \medskip
  \caption{Considered islands of tractability.}
   \label{tab:classes}
 \end{threeparttable}
 \end{table}
}


\subsection{Weak Backdoor Sets} 
If we are given a weak $\CCC$\hy backdoor set of $F$ of size $k$,
we know that $F$ is satisfiable, and we can verify the satisfiability of $F$ by
checking whether at least one of the $2^k$ assignments to the backdoor variables
leads to a formula that belongs to $\CCC$ and is satisfiable.
If the base class allows to find an actual satisfying assignment in polynomial time,
as is usually the case, we can find a satisfying assignment of $F$ in $2^k n^{O(1)}$ time.
Can we find such a backdoor set quickly if it exists?
For all reasonable base classes
$\CCC$ it is NP-hard to decide, given a CNF formula $F$ and an integer
$k$, whether $F$ has a strong or weak $\CCC$\hy backdoor set of size at
most $k$. On the other hand, the problem is clearly solvable in time
$n^{k+O(1)}$. The question is
whether we can get $k$ out of the exponent, and find a backdoor set in
time $f(k)n^{O(1)}$, i.e., is weak backdoor set detection
\emph{fixed-parameter tractable (FPT)} in $k$?
Over the last couple of years, this
question has been answered for various base classes $\CCC$;
Table~\ref{tab:strong} gives an overview of some of the known results.

For general CNF, the detection of weak
$\CCC$\hy backdoor sets is $\W[2]$\hy hard for all reasonable base
classes $\CCC$. For some base classes the problem becomes
\FPT\ if clause lengths are bounded.
All \longshort{fixed-parameter tractability}{\FPT} results for weak backdoor set detection in Table~\ref{tab:weak} are due
to the fact that for $r$\hy CNF formulas, where $r\ge 3$ is a fixed constant, membership in the considered base
class can be characterized by certain obstructions of bounded size. Formally,
say that a base class $\cC$ has the \emph{small obstruction property} if there is a family $\cF$
of CNF formulas, each with a finite number of clauses, such that for any CNF formula $F$,
$F\in \cC$ \myiff $F$ contains no subset of clauses isomorphic to a formula in $\cF$.
Hence, if a
base class $\CCC$ has this property, fixed-parameter tractability for weak $\CCC$\hy
backdoor set detection for $r$\hy CNF formulas can be established by a bounded search tree algorithm.

The base class $\FOREST$ is another class for which the detection of
weak backdoor sets is \W[2]-hard for general CNF formulas (Theorem \ref{thm:Whard}).
For $r$\hy CNF formulas the above argument does not apply because
$\FOREST$ does not have the small obstruction property.
Nevertheless, we can still show that the weak \Forest backdoor set detection problem is \longshort{\fpt}{\FPT} for $r$\hy CNF formulas,
for any fixed $r\ge 3$ (Theorem \ref{thm:weak}). This is our first main result.

\subsection{Strong Backdoor Sets} 
Given a strong $\CCC$\hy backdoor set of size $k$ of a formula
$F$, one can decide whether $F$ is satisfiable by
$2^k$ polynomial checks. 
In Table \ref{tab:strong},
$\HORN$ and $\TWOCNF$ are the only base classes for
which strong backdoor set detection is \FPT\ in
general. A possible reason for the special status of these two classes is
the fact that they have the \emph{deletion property}: for $\CCC\in
\{\HORN,\TWOCNF\}$ a set $X$ of variables is a strong $\CCC$\hy backdoor
set of \longversion{a CNF formula }$F$ \myiff $X$ is a \emph{deletion $\CCC$\hy backdoor set} of $F$,
i.e., the formula \longversion{$F-X$, }obtained from $F$ by deleting all positive and
negative occurrences of the variables in $X$, is in $\CCC$. The advantage of the
deletion property is that it simplifies the search for a strong backdoor
set. Its disadvantage is that the backdoor set
cannot ``repair'' \longshort{the given formula $F$}{a formula} differently for different truth
assignments of the backdoor variables, and thus it does not use the full
power of all the partial assignments. Indeed, for other base
classes one can construct formulas with small strong backdoor sets
whose smallest deletion backdoor sets are arbitrarily large. In view of these results, one
wonders whether a small strong backdoor set can be found efficiently
for a base class that does not have the deletion property.
Our second main result provides a positive answer.
Namely we exhibit an \FPT\ algorithm, which, for a CNF formula $F$ and a
positive integer parameter $k$, either concludes that $F$ has no strong \FOREST{}\hy backdoor set
of size \longversion{at most }$k$ or concludes that $F$ has a strong \Forest{}\hy backdoor set of size at most $2^k$ (Theorem~\ref{thm:strong}).


\begin{table}[tb]
\newcommand{\scite}[1]{${}^\text{{\protect\cite{#1}}}$}
\newcommand{\both}[1]{\multicolumn{2}{c}{#1}}
\begin{tabular*}{\textwidth}{@{\extracolsep{\fill}} *{5}{l} }
  \toprule
    & \both{\emph{Weak}}    & \both{\emph{Strong}}\\
  \cline{2-3}\cline{4-5}
  \emph{Base Class}      & \textsc{CNF} & \textsc{$r$\hy CNF}  & \textsc{CNF} & \textsc{$r$\hy CNF} \\
  \midrule
  $\HORN$  & $\W[2]$-h \scite{NishimuraRagdeSzeider04-informal}   
           & FPT
           & FPT \scite{NishimuraRagdeSzeider04-informal}   
           & FPT \scite{NishimuraRagdeSzeider04-informal}\\
  $\TWOCNF$& $\W[2]$-h \scite{NishimuraRagdeSzeider04-informal}   
           & FPT
           & FPT \scite{NishimuraRagdeSzeider04-informal}   
           & FPT \scite{NishimuraRagdeSzeider04-informal}   \\
  $\UP$    & W[P]-c \scite{Szeider05d} 
           & W[P]-c \scite{Szeider05d}
           & W[P]-c \scite{Szeider05d} 
           & W[P]-c \scite{Szeider05d}\\ 
  $\RHORN$  
  & $\W[2]$-h \scite{GaspersSzeider11festschrift}
  & $\W[2]$-h \scite{GaspersSzeider11festschrift}
  & $\W[2]$-h  \scite{GaspersSzeider11festschrift}
  & open    \\
  $\CLU$  
  & W[2]-h \scite{NishimuraRagdeSzeider07} 
  & FPT
  & W[2]-h \scite{NishimuraRagdeSzeider07} 
  & FPT \scite{NishimuraRagdeSzeider07} \\
   \bottomrule
\end{tabular*}
 \medskip
 \caption{The parameterized complexity of finding weak and strong backdoor
   sets of CNF formulas and $r$\hy CNF formulas, where $r\ge 3$ is a fixed integer.
   See \cite{GaspersSzeider11festschrift} for a survey.}
  \label{tab:bds}\label{tab:weak}\label{tab:strong}
\end{table}

This \FPT-approximation result is interesting for several reasons.
First, it implies that SAT and \#SAT are \FPT,
parameterized by the size of a smallest strong \FOREST{}\hy backdoor set.
Second, (unlike the size of a smallest deletion \FOREST{}\hy backdoor set) the size of a smallest strong \FOREST{}\hy backdoor set
is incomparable to the treewidth of the incidence graph. Hence the result applies to
formulas that cannot be solved efficiently by other known methods.
Finally, it exemplifies a base class that does not satisfy the
deletion property, for which strong backdoor sets are FPT-approximable.

\subsection{\#SAT and Implied Cycle Cutsets}

Our second main result, Theorem \ref{thm:strong}, has applications to
the model counting problem \#SAT, a problem that occurs, for instance,
in the context of Bayesian
Reasoning~\cite{BacchusDalmaoPitassi03,Roth96}.  \#SAT is
\#P-complete~\cite{Valiant79b} and remains \#P-hard even for monotone
2\hy CNF formulas and Horn 2\hy CNF formulas, and it is $\NP$\hy hard to
approximate the number of models of a formula with $n$ variables within
$2^{n^{1-\epsilon}}$ for $\epsilon>0$, even for
monotone 2\hy CNF formulas and Horn 2\hy CNF
formulas~\cite{Roth96}. A common approach to solve \#SAT is to find a
small \emph{cycle cutset} (or feedback vertex set) of variables of the
given CNF formula, and by summing up the number of satisfying
assignments of all the acyclic instances one gets by setting the cutset
variables in all possible ways~\cite{Dechter03}.  Such a cycle
cutset is nothing but a deletion \FOREST{}\hy backdoor set. By
considering strong \FOREST{}\hy backdoor sets instead, one can get
super-exponentially smaller sets of variables, and hence a more
powerful method. A strong \FOREST{}\hy backdoor set can be considered as
a an \emph{implied cycle cutset} as it can cut cycles by removing
clauses that are satisfied by  certain truth assignments to the
backdoor variables. Theorem \ref{thm:strong} states that we can find a
small implied cycle cutset efficiently if one exists.

\section{Preliminaries}

\longversion{
 \paragraph{Parameterized Complexity}
 Parameterized Complexity \cite{DowneyFellows99,FlumGrohe06,Niedermeier06} is a two-dimensional framework to classify the complexity of problems based on their
 input size $n$ and some additional parameter $k$.
 It distinguishes between running times of the form $f(k) n^{g(k)}$ where the degree
 of the polynomial depends on $k$ and running times of the form $f(k) n^{O(1)}$ where the exponential part of the running time is independent of $n$.
 The fundamental hierarchy of parameterized complexity classes is
 \begin{align*}
  \FPT \subseteq \W[1] \subseteq \W[2] \cdots \subseteq \XP.
 \end{align*}
 An algorithm for a parameterized problem is an \FPT\ algorithm if there is a function $f$ such that the running time of the algorithm is
 upper bounded by $f(k) n^{O(1)}$.
 A parameterized problem is in \FPT\ (fixed-parameter tractable) if it has an \FPT\ algorithm,
 a problem is in \XP\ if there are functions $f,g$ such that the problem can be solved in time $f(k) n^{g(k)}$,
 and $\W[t]$, $t\ge 1$, are parameterized intractability classes giving strong evidence that a parameterized problem that is hard for any of these classes is not in \FPT.
 These classes are closed under parameterized reductions, which are $f(k) n^{O(1)}$ time reductions where the target parameter is upper bounded by a function of the source parameter.
 All classes in this hierarchy are believed to be distinct. If $\FPT = \W[1]$, then the Exponential Time Hypothesis \cite{ImpagliazzoPaturiZane01} fails \cite{ChenHKX06}.
}
\shortversion{
 We refer to standard textbooks for background in
 parameterized complexity \cite{DowneyFellows99,FlumGrohe06} and graph theory \cite{Diestel10}.
}

\paragraph{Backdoors\shortversion{.}}
A \emph{literal} is a propositional variable $x$ or its negation $\neg x$.
A \emph{clause} is a disjunction of literals that does not contain a complementary
pair $x$ and $\neg x$.
A \emph{propositional formula} in \emph{conjunctive normal form} (CNF
formula) is a conjunction of clauses.
An $r$\hy CNF formula is a CNF formula where each clause contains
at most $r$ literals.\longversion{

}
For a clause $c$, we write $\lit(c)$ and $\var(c)$ for the sets of literals and variables
occurring in $c$, respectively.
For a CNF formula $F$ we write $\cla(F)$ for its set of clauses,
$\lit(F) = \bigcup_{c\in \cla(F)} \lit(c)$ for its set of literals, and
$\var(F) = \bigcup_{c\in \cla(F)} \var(c)$ for its set of variables.

Let $F$ be a CNF formula and $X\subseteq \var(F)$.
We denote by $2^X$ the set of
all mappings $\tau:X\rightarrow \set{0,1}$, the \emph{truth assignments} on $X$.
%
A truth assignment on $X$
can be extended to 
the literals over $X$
by setting $\tau(\neg x) = 1-\tau(x)$ for all $x\in X$.
Given a truth assignment $\tau \in 2^X$ we define
$F[\tau]$ to be the formula obtained from $F$ by removing all clauses $c$
such that $\tau$ sets a literal of $c$ to 1, and removing the literals set to 0
from all remaining clauses.
\longversion{%
A CNF formula }$F$ is \emph{satisfiable} if there is some $\tau\in
2^{\var(F)}$ with $F[\tau]=\emptyset$. 
SAT is the $\NP$-complete problem of deciding whether a given CNF formula is
satisfiable~\cite{Cook71,Levin73}. \#SAT is the \#P-complete problem of
determining the number of distinct $\tau\in 2^{\var(F)}$ with $F[\tau]=\emptyset$ \cite{Valiant79b}.

\BDSsfirst are defined with respect to a fixed class $\cC$ of CNF
formulas, the \emph{base class}.%
\longversion{
 From a base class we
require the following properties:
\begin{enumerate}
 \item $\cC$ can be recognized in polynomial time,
 \item the satisfiability of formulas in $\cC$ can be decided in polynomial time, and
 \item $\cC$ is closed under isomorphisms (\ie, if two formulas differ only in the names of their
variables, then either both or none belong to $\cC$).
\end{enumerate}
A polynomial time algorithm that determines the satisfiability of any CNF formula from $\cC$
is called a sub-solver~\cite{GomesKautzSabharwalSelman08,WilliamsGomesSelman03}.

}
Let \longshort{$B$ be a set of propositional variables and $F$ be a CNF formula.}{$B\subseteq \var(F)$.}
$B$ is a \emph{strong} \emph{$\cC$-\BDS{}} of $F$ if $F[\tau]\in \cC$ for each $\tau \in 2^B$.
$B$ is a \emph{weak} \emph{$\cC$-\BDS{}} of $F$ if there is an assignment $\tau \in 2^B$ such that $F[\tau]$ is satisfiable and $F[\tau]\in \cC$.
$B$ is a \emph{deletion $\cC$-\BDS{}} of $F$ if $F - B \in \cC$, where $F - B = \set{C \setminus \set{x, \neg x : x\in B} : C \in F}$.

The challenging problem is to find a strong, weak, or deletion
$\cC$-\BDS of size at most $k$ if it exists. This leads to
the following backdoor detection problems for
any base class $\cC$.

\pbDefP{\textsc{Strong $\cC$-\BDS Detection}}
{A CNF formula $F$ and an integer $k\geq 0$.}
{The integer $k$.}
{Does $F$ have a strong $\cC$-backdoor set of size at most $k$?}
The problems \textsc{Weak $\cC$-\BDS Detection} and \textsc{Deletion $\cC$-\BDS Detection} are defined
similarly.

\longversion{
 \paragraph{Graphs}
 Let $G=(V,E)$ be a simple, finite graph.
 Let $S \subseteq V$\longversion{ be a subset of its vertices} and $v\in V$\longversion{ be a vertex}.
 We denote by $G - S$ the graph obtained from $G$ by removing all vertices in $S$ and all edges incident to a vertex in $S$.
 We denote by $G[S]$ the graph $G - (V\setminus S)$.
 The (open) neighborhood of $v$ is $N(v) = \set{u\in V : uv\in E}$, the (open) neighborhood of $S$ is $N(S) = \bigcup_{u\in S}N(u)\setminus S$, and their closed
 neighborhoods are $N[v] = N(v)\cup \set{v}$ and $N[S] = N(S)\cup S$, respectively.
 \longversion{The set }$S$ is a \emph{feedback vertex set} if $G - S$ is acyclic, and $S$ is an independent set if $G[S]$ has no edge.
 \longversion{
 
 A \emph{tree decomposition} of $G$ is a pair 
 $(\{X_i : i\in I\},T)$
 where $X_i \subseteq V$, $i\in I$, and $T$ is a tree with elements
 of $I$ as nodes
 such that:
 \begin{enumerate}
   \item $\bigcup_{i\in I} X_i = V$;
   \item $\forall uv\in E$, $\exists i \in I$ such that $\{u,v\} 
 \subseteq X_i$;
   \item $\forall i,j,k \in I$, if $j$ is on the path from $i$ to $k$ in $T$ 
 then $X_i \cap X_k \subseteq X_j$.
 \end{enumerate}
 The \emph{width} of a tree decomposition is $\max_{i \in I} |X_i|-1$. }%
 The \emph{treewidth} \cite{RobertsonSeymour86} of $G$
 \longversion{is the minimum width taken over all tree decompositions
 of $G$ and it }is denoted by $\tw(G)$.
}

\paragraph{Acyclic Formulas\shortversion{.}}
The \emph{incidence graph} of a CNF formula $F$ is the bipartite graph $\inc(F)=(V,E)$ with
$V = \var(F) \cup \cla(F)$ and for a variable $x \in \var(F)$ and a clause $c \in \cla(F)$
we have $x c \in E$ if $x\in \var(c)$. The edges of $G$ may be annotated by a function
$\sign: E\rightarrow \set{\signp,\signm}$. The \emph{sign} of an edge $x c$ is
\begin{align*}
\sign(xc) = 
\begin{cases}
\signp & \text{if } x \in \lit(c), \text{ and}\\
\signm & \text{if } \neg x \in \lit(c)\enspace.
\end{cases}
\end{align*}
A \emph{cycle} in $F$ is a cycle in $\inc(F)$.
The formula $F$ is \emph{\acyclic{}} if $\inc(F)$ is acyclic \cite{Fagin83}.
We denote by \Forest the set of all \acyclic CNF formulas.

The satisfiability of formulas from \Forest can be decided in polynomial time,
and even the number of satisfying assignments of formulas from \Forest can be determined in polynomial time
\cite{FischerMakowskyRavve06,SamerSzeider10}.

The \emph{strong clause-literal graph} of  $F$ is the graph $\slit(F)=(V,E)$ with
$V = \lit(F) \cup \cla(F)$.
There is an edge $u c\in E$, with $u\in \lit(F)$ and $c\in \cla(F)$ if $u \in \lit(c)$
and there is an edge $u v\in E$, with $u,v\in \lit(F)$ if $u= \neg v$ or $\neg u=v$.%
\longversion{
 The following lemma clarifies the relation of the strong clause-literal graph with \FBDSs.
}

\begin{restatable}{lemma}{LemSlit}\label{lem:slit}\shortversion{\textup{($\star$)}}
 Let $F$ be a CNF formula, $\tau$ be an assignment to $B\subseteq \var(F)$.
 The formula $F[\tau]$ is acyclic \myiff $\textup{\slit}(F) - N[\truef(\tau)]$ is acyclic.
\end{restatable}
\longversion{\input{proofLemSlit}}

\noindent
It follows that there is a bijection between assignments $\tau$ such that $F[\tau]$ is acyclic and
independent sets $Y \subseteq \lit(F)$ in $\slit(F)$ such that $\slit(F) - N[Y]$ is acyclic.

\section{Background and Methods}

The simplest type of \FBDSs are deletion \FBDSs. In the incidence graph, they correspond to \fvss that
are subsets of $\var(F)$. Therefore, algorithms solving slight generalizations of \textsc{Feedback Vertex Set}
can be used to solve the \DBDSpb problem. By results from \cite{ChenFLLV08} and \cite{FominGPR08},
\DBDSpb is \FPT\ and can be solved in time $5^k \cdot \CCard{F}^{O(1)}$ and in time $1.7548^n \cdot \CCard{F}^{O(1)}$,
where $n$ is the number of variables of $F$ and $\CCard{F} = \sum_{c\in \cla(F)} |\lit(c)|$ denotes the formula length.

Any deletion \FBDS $B$ of a CNF formula $F$ is also a strong \FBDS of $F$ and if $F$ is satisfiable, then $B$ is also a weak \FBDS.
In recent years SAT has been studied with respect to several width parameters of its primal, dual, and incidence graph
\cite{AlekhnovichRazborov02,FischerMakowskyRavve06,GanianHlinenyObdrzalek10,OrdyniakPaulusmaSzeider10,SamerSzeider10,Szeider04b}.
Several such parameters are more general than the size of a smallest deletion \FBDS, such as the treewidth or the cliquewidth of the incidence graph.%
\longversion{ Parameterized by the treewidth of the incidence graph SAT is fixed-parameter tractable \cite{FischerMakowskyRavve06,SamerSzeider10},
but the parameterization by cliquewidth is \W[1]-hard, even when an optimal cliquewidth expression is provided \cite{OrdyniakPaulusmaSzeider10}.
Parameterized by the cliquewidth of the \emph{directed} incidence graph (the orientation of an edge indicates whether the variable occurs positively or negatively),
SAT becomes \fpt \cite{FischerMakowskyRavve06,GanianHlinenyObdrzalek10}. It is not known whether the problem of computing an optimal cliquewidth expression of a directed
graph is \FPT\ parameterized by the cliquewidth, but it has an \FPT\ approximation algorithm \cite{KanteR11}, which is sufficient to state that SAT is \FPT\ parameterized by
the cliquewidth of the directed incidence graph.}

The size of a smallest weak and strong \FBDS is incomparable to treewidth and clique\-width.
On one hand, one can construct formulas with arbitrary large \FBDSs by taking the disjoint union of formulas with bounded width.
On the other hand, consider an $r\times r$ grid of variables and subdivide each edge by a clause. 
Now, add a variable $x$ that is contained positively in all clauses subdividing horizontal edges and negatively in all other clauses.
The set $\set{x}$ is a weak and strong \FBDS of this formula, but the treewidth and cliquewidth of the formula depend on $r$.
Therefore, weak and strong \FBDSs have the potential of augmenting the tractable fragments of SAT formulas.%
\longversion{

It would be tempting to use Chen \etal's \FPT\ algorithm for \textsc{Directed Feedback Vertex Set} \cite{ChenLiuLuOsullivanRazgon08} for the detection of deletion \BDSs.
The corresponding base class would contain all CNF formulas with acyclic directed incidence graphs.
Unfortunately this class is not suited as a base class since it contains formulas where each clause contains either only positive literals or only negative literals,
and SAT is well known to be NP-hard for such formulas~\cite{GareyJohnson79}.
}

\medskip
In the remainder of this section we outline our algorithms.
To find a weak or strong \FBDS, consider the incidence graph $G=\inc(F)$ of the input formula $F$.
By Robertson and Seymour's Grid Minor Theorem \cite{RobertsonSeymour86b} there is a function $f : \mathbb{N}\rightarrow\mathbb{N}$
such that for every integer $r$, either $\tw(G)\le f(r)$ or $G$ has an $r \times r$ grid minor. \shortversion{Here, $\tw(G)$ denotes the treewidth of $G$. }%
Choosing $r$ to be a function of the parameter $k$, it suffices to solve the problems for incidence graphs whose treewidth
is upper bounded by a function of $k$, and for incidence graphs that contain an $r \times r$ grid minor, where $r$ is lower
bounded by a function of $k$. The former case can be solved by invoking Courcelle's theorem \cite{Courcelle90} as the \FBDS
\textsc{Detection} problems can be defined in Monadic Second Order Logic. In the latter case we make use of
the fact that $G$ contains many \vdcs and we consider several cases how these cycles might disappear from
$\inc(F)$ by assigning values to variables.

In order to obtain slightly better bounds, instead of relying on the Grid Minor Theorem we use the Erd\H{o}s-P\'{o}sa Theorem \cite{ErdosPosa65} and an algorithmization
by Bodlaender \cite{Bodlaender94} to distinguish between the cases where $G$ has small treewidth (in fact, a small \fvs) or many \vdcs.

\begin{theorem}[\cite{ErdosPosa65}]
Let $k\ge 0$ be an integer. There exists a function $f(k) = O(k \log k)$ such that every graph either contains
$k$ \vdcs or has a \fvs of size $f(k)$.
\end{theorem}


\begin{theorem}[\cite{Bodlaender94}]\label{thm:Bodlaender}
Let $k\ge 2$ be an integer. There exists an $O(n)$ time algorithm, taking as input a graph $G$ on $n$ vertices, that either finds $k$ \vdcs
in $G$ or finds a \fvs of $G$ of size at most $12 k^2 -27 k + 15$.
\end{theorem}

\noindent
We will use Theorem \ref{thm:Bodlaender} to distinguish between
the case where $G$ has a \fvs of size $\kfvs$ and
the case where $G$ has $\kcycles$ \vdcs, for some function $\fcycles:\mathbb{N}\rightarrow\mathbb{N}$,
where $\kfvs = 12 (\kcycles)^2 -27 \kcycles + 15$.

Suppose $G$ has a \fvs $W$ of size $\kfvs$.
By adding $W$ to every bag of an optimal
tree decomposition of $G - W$, we obtain a tree decomposition of $G$ of width at most $\kfvs+1$.
\longversion{%
We use Courcelle's theorem \cite{Courcelle90}, stating that every problem that can be defined in Monadic Second Order Logic (MSO)
can be solved in linear time on structures of bounded treewidth. We use the notation of \cite{FlumGrohe06}.

\pbDefP{tw-MSO}
{A relational structure $\mathcal{A}$ and an MSO-sentence $\varphi$.}
{$\tw(A)+|\varphi|$.}
{Decide whether $\mathcal{A} \models \varphi$.}

\begin{theorem}[\cite{Courcelle90}]\label{thm:Courcelle}
 tw-MSO is \fpt.
\end{theorem}

\noindent
In Lemmas \ref{lem:weakfvs} and \ref{lem:strongfvs} we will define the \textsc{Weak} and \SBDSpb problems
as MSO-sentences, and Theorem \ref{thm:Courcelle} can then be used to solve the problems when a \fvs of size $\kfvs$
is part of the input.
}\shortversion{%
We then define the \textsc{Weak} and \SBDSpb problems
in Monadic Second Order Logic (MSO) and use Courcelle's theorem \cite{Courcelle90} to conclude.
}

\medskip

Our main arguments come into play when Bodlaender's algorithm returns a set $\cC$ of $\kcycles$ \vdcs of $G$.
The algorithms will then compute a set $S^* \subseteq \var(F)$ whose size is upper bounded by a function of $k$
such that every weak/strong \FBDS of size at most $k$ contains a variable from $S^*$. A standard branching argument will then
be used to recurse. In the case of \WBDSpb, $F$ has a \wBDS of size at most $k$ \myiff there is a variable $x\in S^*$, such that $F[x=0]$ or
$F[x=1]$ has a \wBDS of size at most $k-1$. In the case of \SBDSpb, $F$ has no \sBDS of size at most $k$ if for every variable $x\in S^*$,
$F[x=0]$ or $F[x=1]$ has no \sBDS of size at most $k-1$, and if $F[x=0]$ and $F[x=1]$ have \sBDSs $B$ and $B'$ of size at most $2^{k-1}-1$,
then $B\cup B'\cup \set{x}$ is a \sBDS of $F$ of size at most $2^k-1$, leading to a factor $2^k/k$ approximation.

In order to compute the set $S^*$, the algorithms consider how the cycles in $\cC$ can interact with a \BDS.
Let $x$ be a variable and $C$ a cycle in $G$.
In the case of \wBDSs, we say that $x$ \emph{kills}\footnote{We apologize for the violent language.} $C$ if either $\inc(F[x = \true])$ or $\inc(F[x = \false])$ does not contain $C$.
In the case of \sBDSs, we say that $x$ \emph{kills} $C$ if neither $\inc(F[x = \true])$ nor $\inc(F[x = \false])$ contain $C$.
We say that $x$ kills $C$ \emph{internally} if $x\in C$, and that $x$ kills $C$ \emph{externally} if $x$ kills $C$ but does not kill it internally.
In any \FBDS of size at most $k$, at most $k$ cycles from $\cC$ can be killed internally, since all cycles from $\cC$ are vertex-disjoint. The algorithms
go over all possible choices of selecting $k$ cycles from $\cC$ that may be killed internally. All other cycles $\cC'$ need to be killed
externally. The algorithms now aim at computing a set $S$ such that any weak/strong \FBDS of size at most $k$ which is a subset of
$\var(F) \setminus \bigcup_{C\in \cC'} \var(C)$ contains a variable from $S$.
Computing the set $S$ is the most challenging part of this work.
In the algorithm for \wBDSs there is an intricate interplay between several cases, making use of bounded clause lengths.
In the algorithm for \sBDSs a further argument is needed to obtain a more structured interaction between the considered cycles and their external killers.

\section{\WBDSs}
\label{sec:weak}

By a parameterized reduction from \textsc{Hitting Set}, \WBDSpb is easily shown to be \W[2]-hard.

\begin{restatable}{theorem}{ThmWhard}\label{thm:Whard}\shortversion{\textup{($\star$)}}
\WBDSpb is \W[2]-hard.
\end{restatable}
\longversion{\input{proofThmWhard}}

\noindent
In the remainder of this section, we consider the \WBDSpb problem for $r$\hy CNF formulas, for any fixed integer $r\ge 3$.
Let $F$ be an $r$\hy CNF formula, and consider its incidence graph $G=(V,E)=\inc(F)$.
We use Theorem \ref{thm:Bodlaender} to distinguish between the case where $G$ has
many \vdcs and the case where $G$ has a small \fvs. If $G$ has a small
\fvs, the problem is expressed in MSO and solved by Courcelle's theorem.

\begin{restatable}{lemma}{LemWeakfvs}\label{lem:weakfvs}\shortversion{\textup{($\star$)}}
Given a \fvs of $\inc(F)$ of size $\kfvs$, \WBDSpb is \fpt.
\end{restatable}
\longversion{\input{proofLemWeakfvs}}

\noindent
Let $\cC=\set{C_1, \dots, C_{\kcycles}}$ denote \vdcs in $G$, with $\kcycles = 2 \kcur +1$. 
We describe an algorithm that
finds a set $S^*$ of $O(r4^k \kcur^6)$ variables from $\var(F)$ such that any \wBDS of $F$
of size at most $\kcur$ contains a variable from $S^*$.

We will use several functions of $k$ in our arguments. Let
\longshort{%
	\begin{align*}
	\kextcycles  &:= \kcycles-\kcur,\\
	\kmulti      &:= 4 \kcur,\\
	\ksupp       &:= (r-3)\cdot (\kcur^3+9) + 4 \kcur^2 + \kcur, \text{ and}\\
	\koverlap    &:= (r-2)\cdot (\kcur \cdot \kmulti)^2+\kcur \enspace.
	\end{align*}
}{
	\begin{align*}
	\kextcycles  &:= \kcycles-\kcur, & \ksupp       &:= (r-3)\cdot (\kcur^3+9) + 4 \kcur^2 + \kcur,\\
	\kmulti      &:= 4 \kcur, \text{ and} & \koverlap    &:= (r-2)\cdot (\kcur \cdot \kmulti)^2+\kcur \enspace.
	\end{align*}
}
\noindent
Let $C$ be a cycle in $G$ and $x\in \var(F)$.
Recall that $x$ \emph{kills} $C$ \emph{internally} if $x\in C$.
In this case, $x$ is an \emph{internal killer} for $C$.
We say that $x$ \emph{kills} $C$ \emph{externally} if $x\notin C$ and there is a clause $u \in \cla(F) \cap C$ such that $x u\in E$.
In this case, $x$ is an \emph{external killer} for $C$.
%
We first dispense with cycles that are killed internally.
Our algorithm goes through all $\binom{\kcycles}{\kcur}$ ways to choose $\kcur$ cycles from $\cC$ that may be killed internally.
\myWlog, let $C_{\kextcycles+1},\linebreak[1] \dots, \linebreak[1]C_{\kcycles}$ denote the cycles that may be killed internally.
All other cycles $\cC'=\set{C_1, \dots, C_{\kextcycles}}$ need to be killed externally.
Let $\var'(F) = \var(F) \setminus \bigcup_{i=1}^{\kextcycles} \var(C_i)$ denote the variables that may be selected in a \wBDS killing no cycle from $\cC'$ internally. From now on, consider only external killers from $\var'(F)$.
The algorithm will find a set $S$ of $O(rk^6)$ variables such that $S$ contains a variable from any \wBDS $B\subseteq \var'(F)$ of $F$ with $|B|\leq \kcur$.
The algorithm first computes the set of external killers (from $\var'(F)$) for each of these cycles.
Then the algorithm applies the first applicable from the following rules.

\begin{restatable}[No External Killer]{myrule}{RuleNoExtKiller}\label{rule:NoExtKiller}
 If there is a $C_i\in \cC'$ that has no external killer, then set $S := \emptyset$.
\end{restatable}

For each $i\in\set{1,\dots,\kextcycles}$, let $x_i$ be an external killer of $C_i$ that has a maximum number of neighbors in $C_i$.

\begin{restatable}[Multi-Killer Unsupported]{myrule}{RuleMultiKillUnsupp}\label{rule:MultiKillUnsupp}
 If there is a $C_i\in \cC'$ such that $x_i$ has $\ell \ge \kmulti$ neighbors in $C_i$ and at most $\ksupp$ external killers of $C_i$ have
 at least $\ell/(2\kcur)$ neighbors in $C_i$, then include all these external killers in $S$.
\end{restatable}

\begin{restatable}[Multi-Killer Supported]{myrule}{RuleMultiKillSupp}\label{rule:MultiKillSupp}
 If there is a $C_i\in \cC'$ such that $x_i$ has $\ell \ge \kmulti$ neighbors in $C_i$ and more than $\ksupp$ external killers of $C_i$ have
 at least $\ell/(2\kcur)$ neighbors in $C_i$, then set $S := \set{x_i}$.
\end{restatable}

\begin{restatable}[Large Overlap]{myrule}{RuleLargeOverlap}\label{rule:LargeOverlap}
 If there are two cycles $C_i,C_j\in \cC'$, with at least $\koverlap$ common external killers,
 then set $S := \emptyset$.
\end{restatable}

\begin{restatable}[Small Overlap]{myrule}{RuleSmallOverlap}\label{rule:SmallOverlap}
 Include in $S$ all vertices that are common external killers of at least two cycles from $\cC'$.
\end{restatable}

\begin{restatable}{lemma}{LemWeakCorrect}\label{lem:weakcorrect}\shortversion{\textup{($\star$)}}
 Rules \ref{rule:NoExtKiller}--\ref{rule:SmallOverlap} are sound.
\end{restatable}
\longversion{\input{proofLemWeakCorrect}}

\begin{restatable}{lemma}{LemWeakcycles}\label{lem:weakcycles}\shortversion{\textup{($\star$)}}
 There is an \FPT\ algorithm, which, given an $r$\hy CNF formula $F$, a positive integer parameter $\kcur$, and $\kcycles$ \vdcs
 in $\inc(F)$, finds a set $S^*$ of $O(r 4^k k^6)$ variables in $F$ such that
 every \wBDS of $F$ of size at most $\kcur$ contains a variable from $S^*$.
\end{restatable}
\longversion{\input{proofLemWeakcycles}}


\noindent
Our \FPT\ algorithm for \WBDSpb, restricted to $r$\hy CNF formulas, $r\ge 3$, is now easily obtained.

\begin{restatable}{theorem}{ThmWeak}\label{thm:weak}\shortversion{\textup{($\star$)}}
For any fixed $r\ge 3$, \WBDSpb is \fpt for $r$\hy CNF formulas.
\end{restatable}
\longversion{\input{proofThmWeak}}

\section{\SBDSs}
\label{sec:strong}

In this section, we design an algorithm, which, for a CNF formula $F$ and an integer $k$, either concludes that $F$
has no \sBDS of size at most $k$ or concludes that $F$ has a \sBDS of size at most $2^k$.

Let $G=(V,E)=\inc(F)$\longversion{ denote the incidence graph of $F$}.
Again, we consider the cases where $G$ has a small \fvs or a large number of \vdcs separately.
Let
	\begin{align*}
	 \kcycles &= k^2 2^{k-1}+k+1,\\
	 \kextcycles &= \kcycles-k, \text{ and}\\
	 \kfvs &= 12 (\kcycles)^2-27 \kcycles+15\enspace.
	\end{align*}
%
The case where $G$ has a small \fvs is again solved by formulating the problem in
MSO and using Courcelle's theorem.

\begin{restatable}{lemma}{LemStrongfvs}\label{lem:strongfvs}\shortversion{\textup{($\star$)}}
Given a \fvs of $\inc(F)$ of size $\kfvs$, \SBDSpb is \fpt.
\end{restatable}
\longversion{\input{proofLemStrongfvs}}

\noindent
Let $\cC = \set{C_1, \dots, C_{\kcycles}}$ denote \vdcs in $G$. We refer to these cycles as $\cC$-cycles.
The aim is to compute a
set $S^*\subseteq \var(F)$ of size $O(k^{2k}2^{k^2-k})$ such that every \sBDS of $F$ of size at most $k$ contains a variable from $S^*$.

Let $C$ be a cycle in $G$ and $x\in \var(F)$.
Recall that $x$ \emph{kills} $C$ \emph{internally} if $x\in C$.
In this case, $x$ is an \emph{internal killer} for $C$.
We say that $x$ \emph{kills} $C$ \emph{externally} if $x \notin C$ and there are two clauses $u, v \in \cla(F) \cap C$ such that
$x \in \lit(u)$ and $\neg x \in \lit(v)$.
In this case, $x$ is an \emph{external killer} for $C$ and $x$ kills $C$ externally \emph{in} $u$ and $v$.
%
As described earlier,
our algorithm goes through all $\binom{\kcycles}{k}$ ways to choose $k$ $\cC$-cycles
that may be killed internally. \myWlog, let $C_{\kextcycles+1}, \dots, C_\kcycles$ denote the cycles that may be
killed internally.
All other cycles $\cC'=\set{C_1, \dots, C_{\kextcycles}}$ need to be killed externally.
We refer to these cycles as $\cC'$-cycles.
Let $\var'(F) = \var(F) \setminus \bigcup_{i=1}^{\kextcycles} \var(C_i)$ denote the variables that may be selected in a \sBDS killing no $\cC'$-cycle internally. From now on, consider only external killers from $\var'(F)$.
The algorithm will find a set $S$ of at most $2$ variables such that $S$ contains a variable from any \sBDS $B\subseteq \var'(F)$ of $F$ with $|B|\leq \kcur$.
External killers and $\cC'$-cycles might be adjacent in many different ways. The following procedure defines $\cC x$-cycles that
have a much more structured interaction with their external killers.

For each cycle $C_i\in \cC'$
consider vertices $x_i,u_i,v_i$ such that $x_i \in \var'(F)$ kills $C_i$ externally in $u_i$ and $v_i$
 and there is a path $P_i$ from $u_i$ to $v_i$ along the cycle $C_i$ such that if any variable from $\var'(F)$ kills $C_i$ externally in two clauses $u_i'$ and $v_i'$
 such that $u_i',v_i'\in P_i$, then $\set{u_i,v_i} = \set{u_i',v_i'}$.
Let $Cx_i$ denote the cycle $P_i\cup x_i$. We refer to the cycles in $\cC x=\set{Cx_1, \dots, Cx_{\kextcycles}}$ as $\cC x$-cycles.

\begin{observation}
 Every external killer $y$ of a $\cC x$-cycle $Cx_i$ is incident to $u_i$ and $v_i$ and $\sign(y u_i) \neq \sign(y v_i)$.
\end{observation}

\noindent
Indeed, an external killer of $C_i$ that is adjacent to two vertices from $P_i$ with distinct signs is adjacent to $u_i$ and $v_i$.
Moreover, any external killer of $Cx_i$ is a killer for $C_i$ that is adjacent to two vertices from $P_i$ with different signs.
Thus, any external killer of $Cx_i$ is adjacent to $u_i$ and $v_i$.

We will be interested in external killers for $\cC'$-cycles that also kill the corresponding $\cC x$-cycles. That is, we are going to
restrict our attention to vertices in $\var'(F)$ that kill $Cx_i$. An external killer of a $\cC'$-cycle $C_i$ is \emph{interesting}
if it is in $\var'(F)$ and it kills $Cx_i$. As each variable that kills a $\cC x$-cycle $Cx_i$ also kills $C_i$, and each $\cC x$ cycle needs to be killed by a variable
from any \sBDS, we may indeed restrict our attention to interesting external killers of $\cC'$-cycles.

We are now ready to formulate the rules to construct the set $S$ containing at least one variable from any \sBDS $B\subseteq \var'(F)$ of $F$ of size at most $k$.
These rules are applied in the order of their appearance%
\longversion{, which means that a rule is only applicable if all previous rules are not}.

\begin{restatable}[No External Killer]{myrule}{RuleSNoExtKiller}\label{rule:SNoExtKiller}
 If there is a $Cx_i\in \cC x$ such that $Cx_i$ has no external killer, then set $S := \set{x_i}$.
\end{restatable}

\begin{restatable}[Killing Same Cycles]{myrule}{RuleKillingSameCycles}\label{rule:KillingSameCycles}
 If there are vertices $y$ and $z$ and at least $2^{k-1}+1$ $\cC'$-cycles such that both $y$ and $z$ are intersting external killers of
 each of these $\cC'$-cycles,
 then set $S := \set{y,z}$.
\end{restatable}

\begin{restatable}[Killing Many Cycles]{myrule}{RuleKillingManyCycles}\label{rule:KillingManyCycles}
 If there is a $y\in \var'(F)$ such that $y$ is an interesting external killer of at least $k\cdot 2^{k-1}+1$ $\cC'$-cycles, then set $S:=\set{y}$.
\end{restatable}

\begin{restatable}[Too Many Cycles]{myrule}{RuleTooManyCycles}\label{rule:TooManyCycles}
 Set $S:=\emptyset$.
\end{restatable}

\begin{restatable}{lemma}{LemStrongCorrect}\label{lem:strongcorrect}\shortversion{\textup{($\star$)}}
 Rules \ref{rule:SNoExtKiller}--\ref{rule:TooManyCycles} are sound.
\end{restatable}
\longversion{\input{proofLemStrongCorrect}}
\longversion{The following lemma summarizes the construction of the set $S^*$.}

\begin{restatable}{lemma}{LemStrongcycles}\label{lem:strongcycles}\shortversion{\textup{($\star$)}}
There is an \FPT\ algorithm that, given a CNF formula $F$, a positive integer parameter $k$, and $\kcycles$
\vdcs of $\inc(G)$, computes a set $S^*$ of $O(k^{2k} 2^{k^2-k})$ variables from $\var(F)$ such that every \sBDS of $F$ of size at most $k$ includes a variable from $S^*$.
\end{restatable}
\longversion{\input{proofLemStrongcycles}}

\noindent
This can now be used in an \FPT-approximation algorithm for \SBDSpb.
From this algorithm, it follows that SAT and \#SAT, parameterized by the size of a smallest \sBDS, are \longshort{\fpt}{\FPT}.

\begin{restatable}{theorem}{ThmStrong}\label{thm:strong}\shortversion{\textup{($\star$)}}
There is an \FPT\ algorithm, which, for a CNF formula $F$ and a positive integer parameter $k$, either concludes that $F$
has no \sBDS of size at most $k$ or concludes that $F$ has a \sBDS of size at most~$2^k$.
\end{restatable}
\longversion{\input{proofThmStrong}}

\section{Conclusion}

To identify large tractable subproblems the deletion of \fvss has been used in several other contexts,
where instantiations of a smaller number of variables could already lead to acyclic subproblems.
Examples include Nonmonotonic Reasoning \cite{GottlobScarcelloSideri02},
Bayesian inference \cite{BeckerBaryehudaGeiger00,Pearl88}, and
QBF satisfiability \cite{Chen04}. 
We believe that elements from our algorithms and proofs could be used in the design of parameterized, moderately exponential,
and approximation algorithms for \FBDS \textsc{Detection} problems and related problems such as finding a backdoor tree \cite{SamerSzeider08b}
of minimum height or with a minimum number of leaves, for SAT and problems in the above-mentioned contexts.
Indeed, a similar approach has very recently been used to design an \FPT-approximation algorithm for the detection of strong backdoor sets
with respect to the base class of nested CNF formulas~\cite{GaspersSzeider12}.

{\small
\longshort{
 \bibliographystyle{plain}
 \bibliography{literature}
}{
 \bibliographystyle{plain}
 \bibliography{literature-short}
}
}

\shortversion{
\newpage
\appendix

\section{Appendix}

\LemSlit*

\input{proofLemSlit}
\ThmWhard*

\input{proofThmWhard}
\LemWeakfvs*

\input{proofLemWeakfvs}
\LemWeakCorrect*

\input{proofLemWeakCorrect}
\LemWeakcycles*

\input{proofLemWeakcycles}
\ThmWeak*

\input{proofThmWeak}
\LemStrongfvs*

\input{proofLemStrongfvs}
\LemStrongCorrect*

\input{proofLemStrongCorrect}
\LemStrongcycles*

\input{proofLemStrongcycles}
\ThmStrong*

\input{proofThmStrong}
}

\end{document}

%% file: proofLemSlit.tex
\begin{proof}
 There is a one-to-one correspondence between cycles in $\inc(F)$ and cycles in $\slit(F)$.
 Indeed, a cycle $C$ in $\inc(F)$ can be obtained from a cycle $C'$ in $\slit(F)$ by replacing each literal by its variable, and removing a variable $x$ if it is preceded by $x$, and vice-versa.
 The correspondence is one-to-one as no clause contains complementary literals.
 Let $C$ be a cycle in $\inc(F)$ which correponds to the cycle $C'$ in $\slit(F)$.
 We have that $C$ is not a cycle in $\inc(F[\tau])$ if there is a variable $x\in C\cap B$ or there is a clause $c\in C$ and a variable $x\in B$ such that $\tau(x) \in \lit(c)$.
 In the first case, $C'$ is not a cycle in $\slit(F) - N[\truef(\tau)]$ as $\set{x,\neg x} \subseteq N[\truef(\tau)]$.
 In the second case, $C'$ is not a cycle in $\slit(F) - N[\truef(\tau)]$ as $c \in N[\truef(\tau)]$.
 The reverse direction follows similarly.
\myqed \end{proof}

%% file: proofThmWhard.tex
\begin{proof}
 We give a parameterized reduction from the $\W[2]$-complete \textsc{Hitting Set (HS)} problem~\cite{DowneyFellows99}.
 HS has as input a collection $\cS=\set{S_1,\dots,S_m}$ of subsets $S_i$ of a universe $U$ and an integer parameter $k$.
 The question is whether there is a set $Y\subseteq U$ of size $k$ such that every set from $\cS$ contains an element from $Y$.
 In this case, $Y$ is a \emph{hitting set} of $\cS$.

 Create an instance $F$ for \WBDSpb with variables $U \cup \set{z_i,z_i' : S_i\in \cS}$ and for each $S_i\in \cS$, add
 the clauses $c_i=\set{z_i, z_i'}$ and $c_i'=S_i\cup \set{\neg z_i,\neg z_i'}$.
 We claim that $\cS$ has a hitting set of size $k$ \myiff $F$ has a \wBDS of size $k$.
 Let $Y$ be a hitting set of $\cS$ of size $k$. Consider the formula $F' = F[\set{y=1: y\in Y}]$. $F'$
 contains no clause $c_i'$, for any $1\le i\le m$, as $Y$ is a hitting set of $\cS$. Thus, $F'$ contains
 only clauses $c_i$, which are all variable-disjoint. Therefore, $F'$ is acyclic and satisfiable. It follows that $Y$ is a \wBDS for $F$.
 On the other hand, suppose $\tau$ is an assignment to $k$ variables such that $F[\tau]$ is acyclic and satisfiable. Obtain
 $\tau'$ from $\tau$ by replacing each assignment to $z_i$ or $z_i'$ by an assignment setting a literal from $S_i$ to $\true$.
 $F[\tau']$ is also acyclic because any cycle passing through $z_i$, $z_i'$, or $c_i$, also passes through $c_i'$, and
 $c_i'$ is removed from $F[\tau']$. Let $Y=(\tau')^{-1}(1)$. Then each clause $c_i'$ contains a variable from $Y$, otherwise the
 cycle $(c_i',z_i',c_i,z_i,c_i')$ remains. Thus, $Y$ is a hitting set of $\cS$ of size at most $k$.
\myqed \end{proof}

%% file: proofLemWeakfvs.tex
\begin{proof}
For any formula $F$, we define a relational structure $A_F$.
The vocabulary of $A_F$ is $\set{\text{LIT}, \text{CLA}}$, with $\text{LIT} = \lit(F)$ and $\text{CLA} = \cla(F)$.
There is a unary relation $\text{VAR} = \var(F)$, and symmetric binary relations $\text{NEG} = \set{x \neg x: x\in \var(F)}$ and
$\text{EDGE} = \text{NEG} \cup\ \{x c: x\in \text{LIT}, \allowbreak c\in \text{CLA},$ $x\in \lit(c)\}$.

Let $S$ be a \fvs of $\inc(F)$ of size at most $\kfvs$.
A tree decomposition for the graph $\inc(F)$ can be obtained by starting from a trivial tree decomposition of width $1$ for
$\inc(F) - S$ and adding $S$ to every bag of this tree decomposition. A tree decomposition for 
$A_F$ can then be obtained by replacing each vertex by both its literals. This tree decomposition has width at most $2 \kfvs + 3$.

\shortversion{
Courcelle's theorem \cite{Courcelle90}, states that the following problem is \fpt. We use the notation of \cite{FlumGrohe06}.

\pbDefP{tw-MSO}
{A relational structure $\mathcal{A}$ and an MSO-sentence $\varphi$.}
{$\tw(A)+|\varphi|$.}
{Decide whether $\mathcal{A} \models \varphi$.}
}
To determine whether $F$ has a \wBDS of size $k$, we define an MSO-sentence $\varphi(Y)$, checking whether $F[\tau]$ is acyclic, where
$\truef(\tau)=Y$.
Invoking \longshort{Theorem \ref{thm:Courcelle}}{Courcelle's theorem} with the sentence $\exists y_1 \dots \exists y_k (\varphi(\set{y_1,\dots,y_k}))$
will then enable us to find a \wBDS of $F$ of size $k$ if one exists.

Note that $A_F$ encodes the graph $H=\slit(F)$, and by Lemma \ref{lem:slit}
it suffices to find an independent set $Y\subseteq \text{LIT}$ of size $k$ such that $\slit(F) - N[Y]$
is acyclic.

We break up $\varphi$ into several simpler sentences. The following sentence checks whether $Y$ is an assignment.
\begin{align*}
 \varphi_{\text{ass}}(Y) = \forall y (Yy \rightarrow (\text{LIT} y \wedge (\neg \exists z (Yz \wedge \text{NEG} yz))))
\end{align*}
To make sure that $H - N[Y]$ is acyclic, it is sufficient that every subgraph of $H$ with minimum degree at least $2$ has a vertex from $Y$
in its closed neighborhood. The following sentence checks whether the set $C$ induces a subgraph with minimum degree at least $2$.
\begin{align*}
 \varphi_{\text{deg2}}(C) = \forall x (Cx \rightarrow \exists y_1 \exists y_2 (C y_1 \wedge C y_2 \wedge y_1 \neq y_2 \wedge \text{EDGE} x y_1 \wedge \text{EDGE} x y_2))
\end{align*}
The following sentence checks whether $C$ has a vertex from $N[Y]$.
\begin{align*}
 \varphi_{\text{kills}}(Y,C) = \exists x \exists y (Cx \wedge Yy \wedge (x=y \vee \text{EDGE} xy))
\end{align*}
Our final MSO-sentence checks whether $Y$ is an independent set of $\text{LIT}$ such that $H - N[Y]$
is acyclic.
\begin{align*}
 \varphi(Y) = \varphi_{\text{ass}}(Y) \wedge \forall C (\varphi_{\text{deg2}}(C) \rightarrow \varphi_{\text{kills}}(Y,C))
\end{align*}
This proves the lemma.
\myqed \end{proof}

%% file: proofLemWeakCorrect.tex
\begin{proof}
We prove the correctness of Rules \ref{rule:NoExtKiller}--\ref{rule:SmallOverlap} in the order of their appearance.

\RuleNoExtKiller*
If $C_i$ has no external killer (from $\var'(F)$), then $F$ has no \wBDS of size $k$ which is a subset of $\var'(F)$. 

Recall that for each $i\in\set{1,\dots,\kextcycles}$, the variable $x_i$ is an external killer of $C_i$ that has a maximum number of neighbors in $C_i$.

\RuleMultiKillUnsupp*
Consider a natural ordering $a_1, \dots, a_{\ell}$ of the neighbors of $x_i$ in $C_i$; i.e., $a_1, \dots, a_{\ell}$ occur in this order on the cycle.
See Figure \ref{fig:mkunsupp}.
For convenience, let $a_0=a_{\ell}$ and $a_{\ell+1}=a_1$.
Let $P_j$ denote the set of vertices that are encountered when moving on the cycle from $a_j$ to $a_{j+1}$ without passing through $a_{j-1}$.
The \BDSs that do not contain $x_i$ need to kill each of the cycles $P_j \cup \set{x_i}, 1\le j\le \ell,$ externally.
All such \BDSs of size at most $\kcur$ necessarily contain a vertex killing at least $\ell/\kcur$ of these cycles,
and such a vertex is an external killer of $C_i$ with at least $\ell/(2\kcur)$ neighbors in $C_i$.

\input{figWeak}

\RuleMultiKillSupp*
Let $W$ denote the set of external killers of $C_i$ with at least $\ell/(2\kcur)$ neighbors in $C_i$. See Figure \ref{fig:mksupp}.
For the sake of contradiction, assume there exists a \wBDS $B\subseteq \var'(F) \setminus \set{x_i}$ of $F$ of size at most $\kcur$.
As $x_i$ has a maximum number of neighbors in $C_i$, there are at most $\kcur \cdot \ell$ edges connecting a vertex from $B$ to a vertex from $C_i$.
Consider the maximal segments $J_1,\dots,J_{s}$ of $C_i$ that do not contain a vertex adjacent to a vertex from $B$.
By the previous observation, $s \le \kcur \cdot \ell$.
Let $H$ be an auxiliary bipartite graph with bipartition $(J,X)$ of its vertex set, where $J = \set{J_1, \dots, J_{s}}$
and $X = W\setminus B$, and an edge from $J_j \in J$ to $x\in X$ if $x$ is adjacent to a vertex from $J_j$ in $G$.
The graph $H$ is acyclic as any cycle in $H$ could naturally be expanded
into a cycle in $G$ avoiding the neighborhood of all vertices in $B$ by replacing vertices in $J$ by paths in the corresponding segments of $C_i$.
However, by counting the number of edges incident to $X$ in $H$, which is the number of edges from $X$ to $C_i$ minus the number of edges from
$X$ to a neighbor of $B$, we obtain that
\longshort{
	\begin{align*}
	 |E(H)| &\ge \frac{\ell}{2\kcur} \cdot |X| - (r-3) \cdot s & \text{\hspace{-5cm}(as any $u\in N(B)\cap C_i$ has at most $r-3$ neighbors in $X$)}\\
	        &\ge |X| + \left(\frac{\ell}{2\kcur}-1\right) \cdot \left( (r-3)(\kcur^3+9)+4k^2 \right) - (r-3) \cdot s\\ && \text{\hspace{-10cm}(as $|X|\ge \ksupp-k=(r-3)(k^3+9)+4k^2$)}\\
	        &= |X| + 2\kcur\ell-4\kcur^2 + (r-3) \left( \left(\frac{\ell}{2\kcur}-1\right) (\kcur^3+9)-s \right)\\
	        &\ge |X| + \kcur\ell + \kcur\cdot (\ell-4\kcur) + (r-3) \left( \ell (\kcur^2/2-\kcur) -\kcur^3+9 \right)\\ && \text{\hspace{-10cm}(as $\ell\ge \kmulti = 4\kcur$ and $s\le \kcur\ell$)}\\
	        &\ge |X| + \kcur\ell + (r-3) \left( \kcur^3-4\kcur^2 +9 \right) & \text{\hspace{-10cm}(as $\ell\ge \kmulti = 4\kcur$)}\\
	        &\ge |X| + \kcur\ell & \text{\hspace{-10cm}(as $r\ge 3$ and $\kcur^3-4\kcur^2 +9 \ge 0$ for any integer $\kcur\ge 1$)}\\
	        &\ge |X| + s & \text{\hspace{0cm}(as $s \le \kcur \cdot \ell$)}\\
	        &= |V(H)|\enspace.
	\end{align*}
}{
	\begin{align*}
		 |E(H)| &\ge \frac{\ell}{2\kcur} \cdot |X| - (r-3) \cdot s\\
		 & & \text{\hspace{-7.5cm}(as any $u\in N(B)\cap C_i$ has at most $r-3$ neighbors in $X$)}\\
		        &\ge |X| + \left(\frac{\ell}{2\kcur}-1\right) \cdot \left( (r-3)(\kcur^3+9)+4k^2 \right) - (r-3) \cdot s\\ && \text{\hspace{-10cm}(as $|X|\ge \ksupp-k=(r-3)(k^3+9)+4k^2$)}\\
		        &= |X| + 2\kcur\ell-4\kcur^2 + (r-3) \left( \left(\frac{\ell}{2\kcur}-1\right) (\kcur^3+9)-s \right)\\
		        &\ge |X| + \kcur\ell + \kcur\cdot (\ell-4\kcur) + (r-3) \left( \ell (\kcur^2/2-\kcur) -\kcur^3+9 \right)\\ && \text{\hspace{-10cm}(as $\ell\ge \kmulti = 4\kcur$ and $s\le \kcur\ell$)}\\
		        &\ge |X| + \kcur\ell + (r-3) \left( \kcur^3-4\kcur^2 +9 \right) & \text{\hspace{-10cm}(as $\ell\ge \kmulti = 4\kcur$)}\\
		        &\ge |X| + \kcur\ell & \text{\hspace{-10cm}(as $r\ge 3$ and $\kcur^3-4\kcur^2 +9 \ge 0$ for any integer $\kcur\ge 1$)}\\
		        &\ge |X| + s & \text{\hspace{0cm}(as $s \le \kcur \cdot \ell$)}\\
		        &= |V(H)|\enspace.
		\end{align*}
}
Thus, $H$ has a cycle, a contradiction.

\RuleLargeOverlap*
Consider any vertex subset $B \subseteq \var'(F)$ of size at most $\kcur$.
By the previous two rules, $|N[B]\cap C_i| \le \kcur \cdot (\kmulti-1)$ and $|N[B]\cap C_j| \le \kcur \cdot (\kmulti-1)$.
We will show that there are two common external killers
$y_1$ and $y_2$ of $C_i$ and $C_j$ such that $G[(\set{y_1,y_2}\cup C_i\cup C_j)\setminus N[B]]$ contains a cycle.
Let us denote $Y$ the set of common external killers of $C_i$ and $C_j$.
As there are at least $\koverlap-k$ edges between vertices from $Y\setminus B$ and vertices from $C_i$, the vertices from $Y\setminus B$ have at least $(\koverlap-\kcur)/(r-2)$ neighbors in $C_i$.
The graph $G - N[B]$ contains at most $\kcur \cdot (\kmulti-1)$ segments of the cycle $C_i$.
There is at least one such segment with at least $\frac{\koverlap-\kcur}{(r-2) \cdot \kcur \cdot (\kmulti-1)} > \kcur \cdot \kmulti$ neighbors in $Y\setminus B$.
At least two of these neighbors, $y_1$ and $y_2$, are adjacent to the same segment of $C_j\setminus N[B]$, creating a cycle in $G[(\set{y_1,y_2}\cup C_i\cup C_j)\setminus N[B]]$.
As $B$ was chosen arbitrarily, $F$ has no \wBDS that is a subset of $\var'(F)$.

\RuleSmallOverlap*
By the pigeonhole principle at least one variable of the \BDS needs to kill at least two cycles from $\cC'=\set{C_1,\dots, C_{\kcur+1}}$ externally.
This vertex is among the common external killers of $C_1,\dots, C_{\kcur+1}$, whose number is upper bounded by
$\frac{(\kcur+1) \cdot \kcur}{2} \cdot (\koverlap-1)$ by the previous rule.
\myqed \end{proof}

%% file: figWeak.tex
\tikzset{var/.style={inner sep=.15em,circle,fill=black,draw},
         clause/.style={minimum size=1mm,rectangle,fill=white,draw},
         label distance=-3pt}

\longshort{
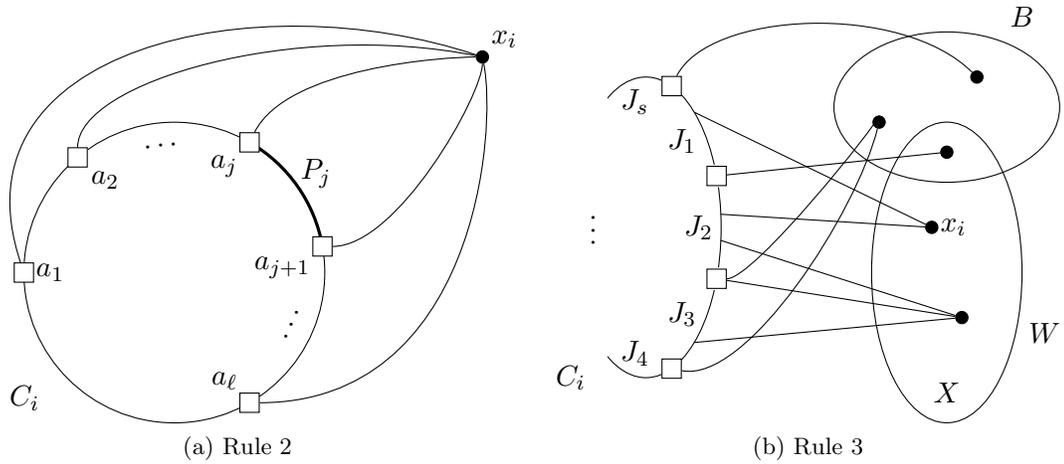
\begin{figure}[tb]
 \centering
 \subfloat[Rule \ref{rule:MultiKillUnsupp}]{\label{fig:mkunsupp}
  \begin{tikzpicture}[xscale=1,yscale=1]
   \draw (0,0) circle (2cm);
   \draw[very thick] +(10:2cm) arc (10:60:2cm);
   \node at +(35:2.3cm) {$P_j$};
   \node at +(220:2.6cm) {$C_i$};
   \node (a1) at +(180:2cm) [clause,label=0:$a_1$] {};
   \node (a2) at +(130:2cm) [clause,label=-50:$a_2$] {};
   \node (a3) at +(60:2cm) [clause,label=-120:$a_j$] {};
   \node (a4) at +(10:2cm) [clause,label=-170:$a_{j+1}$] {};
   \node (a5) at +(-60:2cm) [clause,label=120:$a_{\ell}$] {};
   \node[rotate=5] at +(95:1.7cm) {$\dots$};
   \node[rotate=-115] at +(-25:1.7cm) {$\dots$};
   \node (x) at +(35:5cm) [var,label=35:$x_i$] {};

   \draw (x) .. controls +(-3,1) and +(-1,3) .. (a1);
   \draw (x) .. controls +(-2.5,0.5) and +(0,1) .. (a2);
   \draw (x) .. controls +(-0.5,0) and +(0.5,1) .. (a3);
   \draw (x) .. controls +(0,-0.5) and +(1,0) .. (a4);
   \draw (x) .. controls +(0.2,-1) and +(3,0) .. (a5);
  \end{tikzpicture}
 }
 \subfloat[Rule \ref{rule:MultiKillSupp}]{\label{fig:mksupp}
  \begin{tikzpicture}[xscale=1,yscale=1,x=1cm,y=2cm]
   \begin{scope}
    \clip (-0.5,-1.2) rectangle (1.1,1.2);
    \draw (0,0) ellipse (1cm and 2cm);
   \end{scope}
   \node      at (xyz polar cs:angle=70,radius=1) [var,line width=4pt,draw=white,fill=white] {};
   \node (a1) at (xyz polar cs:angle=70,radius=1) [clause] {};
   \node      at (xyz polar cs:angle=20,radius=1) [var,line width=4pt,draw=white,fill=white] {};
   \node (a2) at (xyz polar cs:angle=20,radius=1) [clause] {};
   \node      at (xyz polar cs:angle=-20,radius=1) [var,line width=4pt,draw=white,fill=white] {};
   \node (a3) at (xyz polar cs:angle=-20,radius=1) [clause] {};
   \node      at (xyz polar cs:angle=-70,radius=1) [var,line width=4pt,draw=white,fill=white] {};
   \node (a4) at (xyz polar cs:angle=-70,radius=1) [clause] {};
   \node      at (xyz polar cs:angle=100,radius=0.85) {$J_s$};
   \node      at (xyz polar cs:angle=50,radius=0.75) {$J_1$};
   \node      at (xyz polar cs:angle=0,radius=0.7) {$J_2$};
   \node      at (xyz polar cs:angle=-50,radius=0.75) {$J_3$};
   \node      at (xyz polar cs:angle=-100,radius=0.85) {$J_4$};
   \node[rotate=90] at (xyz polar cs:angle=180,radius=0.7) {$\dots$};
   \node at (-1,-1) {$C_i$};

   \draw (4,0.8) ellipse (1.5cm and 1cm);
   \node at (5,1.4) {$B$};
   \draw (4,-0.3) ellipse (1cm and 2cm);
   \node at (5.3,-0.7) {$W$};
   \node at (4,-1.1) {$X$};
   \node (x) at (3.8,0) [var,label=right:$x_i$] {};
   \node (y) at (4.2,-0.6) [var] {};
   \node (b1) at (4,0.5) [var] {};
   \node (b2) at (3.1,0.7) [var] {};
   \node (b3) at (4.4,1) [var] {};

   \draw (b3) .. controls +(-1,0.5) and +(0.5,0.5) .. (a1);
   \draw (b1)--(a2);
   \draw (b2) .. controls +(0,0) and +(0.5,0) .. (a3);
   \draw (b2) .. controls +(-0.3,-0.6) and +(1,-0.1) .. (a4);
   \draw (a3)--(y)--(xyz polar cs:angle=-50,radius=1);
   \draw (xyz polar cs:angle=5,radius=1)--(x)--(xyz polar cs:angle=50,radius=1);
   \draw (y)--(xyz polar cs:angle=-5,radius=1);
  \end{tikzpicture}
 }
 \caption{Helper figures for Rules \ref{rule:MultiKillUnsupp} and \ref{rule:MultiKillSupp}. Clauses are represented by squares and variables by solid circles.}
 \label{fig:mk}
\end{figure}
}{
\begin{figure}[tb]
 \centering
 \subfloat[Rule \ref{rule:MultiKillUnsupp}]{\label{fig:mkunsupp}
  \begin{tikzpicture}[xscale=0.8,yscale=0.8]
   \draw (0,0) circle (2cm);
   \draw[very thick] +(10:2cm) arc (10:60:2cm);
   \node at +(35:2.3cm) {$P_j$};
   \node at +(220:2.6cm) {$C_i$};
   \node (a1) at +(180:2cm) [clause,label=0:$a_1$] {};
   \node (a2) at +(130:2cm) [clause,label=-50:$a_2$] {};
   \node (a3) at +(60:2cm) [clause,label=-120:$a_j$] {};
   \node (a4) at +(10:2cm) [clause,label=-170:$a_{j+1}$] {};
   \node (a5) at +(-60:2cm) [clause,label=120:$a_{\ell}$] {};
   \node[rotate=5] at +(95:1.7cm) {$\dots$};
   \node[rotate=-115] at +(-25:1.7cm) {$\dots$};
   \node (x) at +(35:5cm) [var,label=35:$x_i$] {};

   \draw (x) .. controls +(-3,1) and +(-1,3) .. (a1);
   \draw (x) .. controls +(-2.5,0.5) and +(0,1) .. (a2);
   \draw (x) .. controls +(-0.5,0) and +(0.5,1) .. (a3);
   \draw (x) .. controls +(0,-0.5) and +(1,0) .. (a4);
   \draw (x) .. controls +(0.2,-1) and +(3,0) .. (a5);
  \end{tikzpicture}
 }
 \subfloat[Rule \ref{rule:MultiKillSupp}]{\label{fig:mksupp}
  \begin{tikzpicture}[xscale=0.8,yscale=0.7,x=1cm,y=2cm]
   \begin{scope}
    \clip (-0.5,-1.2) rectangle (1.1,1.2);
    \draw (0,0) ellipse (1cm and 2cm);
   \end{scope}
   \node      at (xyz polar cs:angle=70,radius=1) [var,line width=4pt,draw=white,fill=white] {};
   \node (a1) at (xyz polar cs:angle=70,radius=1) [clause] {};
   \node      at (xyz polar cs:angle=20,radius=1) [var,line width=4pt,draw=white,fill=white] {};
   \node (a2) at (xyz polar cs:angle=20,radius=1) [clause] {};
   \node      at (xyz polar cs:angle=-20,radius=1) [var,line width=4pt,draw=white,fill=white] {};
   \node (a3) at (xyz polar cs:angle=-20,radius=1) [clause] {};
   \node      at (xyz polar cs:angle=-70,radius=1) [var,line width=4pt,draw=white,fill=white] {};
   \node (a4) at (xyz polar cs:angle=-70,radius=1) [clause] {};
   \node      at (xyz polar cs:angle=100,radius=0.85) {$J_s$};
   \node      at (xyz polar cs:angle=50,radius=0.75) {$J_1$};
   \node      at (xyz polar cs:angle=0,radius=0.7) {$J_2$};
   \node      at (xyz polar cs:angle=-50,radius=0.75) {$J_3$};
   \node      at (xyz polar cs:angle=-100,radius=0.85) {$J_4$};
   \node[rotate=90] at (xyz polar cs:angle=180,radius=0.7) {$\dots$};
   \node at (-1,-1) {$C_i$};

   \draw (4,0.8) ellipse (1.5cm and 1cm);
   \node at (5,1.4) {$B$};
   \draw (4,-0.3) ellipse (1cm and 2cm);
   \node at (5.3,-0.7) {$W$};
   \node at (4,-1.1) {$X$};
   \node (x) at (3.8,0) [var,label=right:$x_i$] {};
   \node (y) at (4.2,-0.6) [var] {};
   \node (b1) at (4,0.5) [var] {};
   \node (b2) at (3.1,0.7) [var] {};
   \node (b3) at (4.4,1) [var] {};

   \draw (b3) .. controls +(-1,0.5) and +(0.5,0.5) .. (a1);
   \draw (b1)--(a2);
   \draw (b2) .. controls +(0,0) and +(0.5,0) .. (a3);
   \draw (b2) .. controls +(-0.3,-0.6) and +(1,-0.1) .. (a4);
   \draw (a3)--(y)--(xyz polar cs:angle=-50,radius=1);
   \draw (xyz polar cs:angle=5,radius=1)--(x)--(xyz polar cs:angle=50,radius=1);
   \draw (y)--(xyz polar cs:angle=-5,radius=1);
  \end{tikzpicture}
 }
 \caption{Helper figures for Rules \ref{rule:MultiKillUnsupp} and \ref{rule:MultiKillSupp}. Clauses are represented by squares and variables by solid circles.}
 \label{fig:mk}
\end{figure}
}

%% file: proofLemWeakcycles.tex
\begin{proof}
 The algorithm starts with $S^*=\emptyset$.
 For each choice $\mathfrak{C}$ among the $\binom{\kcycles}{\kcur}$ cycles to be killed externally, the algorithm executes one of the described rules.
 It computes a set $S$ such that every \wBDS of $F$ of size at most $\kcur$ respecting $\mathfrak{C}$ contains a variable from $S$.
 We set $S^*$ to be the union of all $S$ that are returned over all choices of cycles to be killed externally.
 As any \wBDS respects at least one such choice, $F$ has a \wBDS of size at most $\kcur$ containing at least one variable from $S^*$ if $F$ has a \wBDS of size at most $\kcur$.

 It remains to bound the size of $S^*$. The largest $S$ are returned by Rule \ref{rule:SmallOverlap} and have size at most $O(r \kcur^2 \cdot \koverlap)=O(r k^6)$.
 As $\binom{\kcycles}{\kcur} \le 2^{2k+1}$, the lemma follows.
\myqed \end{proof}

%% file: proofThmWeak.tex
\begin{proof}
The final \WBDSpb algorithm for $r$-CNF formulas is recursive.
Given an $r$-CNF formula $F$ and an integer $k$, it computes the incidence graph $G=(V,E)=\inc(F)$.
Then the algorithm from Theorem \ref{thm:Bodlaender} is invoked with parameter $k'=\kcycles$.
If that algorithm returns a \fvs of size $O(k^2)$, we can conclude by Lemma \ref{lem:weakfvs}.
Otherwise, a set of $\kcycles$ \vdcs is returned. Then, Lemma \ref{lem:weakcycles} is used to compute a set
$S^*$ such that every \wBDS of $F$ of size at most $k$ contains at least
one variable from $S^*$. The algorithm recursively checks whether any formula $F[s=\false]$ or $F[s=\true]$,
with $s\in S^*$, has a \wBDS of size at most $k-1$ and returns \textsf{true} is any such check was successful
and \textsf{false} otherwise.
\myqed \end{proof}

%% file: proofLemStrongfvs.tex
\begin{proof}
We will use \longshort{Theorem \ref{thm:Courcelle}}{Courcelle's theorem} and the relational structure $A_F$, defined in the proof of Lemma \ref{lem:weakfvs}, to solve this problem.
For a set $X=\set{x_1,\dots,x_{k}}$ our MSO-sentence $\varphi(X)$ will decide whether
$X$ is a \sBDS of $F$. It reuses several subformulas from the proof of Lemma \ref{lem:weakfvs}
and checks, for each assignment to $X$, whether the resulting formula is acyclic.
Invoking \longshort{Theorem \ref{thm:Courcelle}}{Courcelle's theorem} with the sentence $\exists x_1 \dots \exists x_k (\varphi(\set{x_1,\dots,x_k}))$
will then enable us to find a \sBDS of $F$ of size $k$ if one exists.

The following sentence checks whether $X$ is a subset of variables.
\begin{align*}
 \varphi_{\text{var}}(X) = \forall x (Xx \rightarrow \text{VAR} x)
\end{align*}
An assignment of $X$ is a subset of $\text{LIT}$ containing no complementary literals such that every selected literal
is a variable from $X$ or its negation, and for every variable $x$ from $X$, $x$ or $\neg x$ is in $Y$.
The following sentence checks whether $Y$ is an assignment of $X$.
\begin{align*}
 \varphi_{\text{ass}}(X,Y) &= \forall y (Yy \rightarrow ((Xy \vee (\exists z (Xz \wedge \text{NEG} yz)))\\
   & \quad \quad \quad \quad \quad \quad \: \wedge (\forall z (Yz \rightarrow \neg \text{NEG} yz))))\\
   & \quad \; \wedge \forall x (Xx \rightarrow (Yx \vee \exists y (Yy \wedge \text{NEG} xy)))
\end{align*}
Our final sentence checks whether $X$ is a set of variables such that each assignment to $X$ kills all cycles in $\inc(F)$.
\begin{align*}
 \phi(X) = \varphi_{\text{var}}(X) \wedge \forall Y (&\varphi_{\text{ass}}(X,Y) \rightarrow (\forall C(\phi_{\text{deg2}}(C) \rightarrow \phi_{\text{kills}}(Y,C)))
\end{align*}
As we can obtain a tree decomposition for $A_F$ of width $2\kfvs+3$ in polynomial time, and the length of $\phi$ is a function of
$k$, the lemma follows by \longshort{Theorem \ref{thm:Courcelle}}{Courcelle's theorem}.
\myqed \end{proof}

%% file: proofLemStrongCorrect.tex
\begin{proof}
We prove the correctness of Rules \ref{rule:SNoExtKiller}--\ref{rule:TooManyCycles} in the order of their appearance.

\RuleSNoExtKiller*
The correctness of Rule \ref{rule:SNoExtKiller} follows since $x_i$ is the only interesting external killer of $C_i$.

\RuleKillingSameCycles*
We will show that at least one of $y$ and $z$ is in any \sBDS $B\subseteq \var'(F)$ of $F$ of size $k$.
Suppose otherwise and consider a \sBDS $B\subseteq \var'(F) \setminus \set{y,z}$ of $F$ of size $k$.
Consider the $\cC'$-cycles for which $y$ and $z$ are interesting external killers and the set $U$ of all variables $u_i,v_i$ of each such $\cC'$-cycle
$C_i$ as defined above. Note that $|U|\ge 2^k+2$.
We iteratively define a truth assignment $\tau$ to $B=\{b_1,\dots,b_k\}$.
Initially, all vertices in $U$ are unmarked.
At iteration $i$, let $U_i$ and $\overline{U}_i$ denote the set of unmarked vertices
from $U$ that are incident with positive and negative edges to $b_i$, respectively.
Set $\tau(b_i)=\true$ if $|\overline{U}_i|\ge |U_i|$,
and set $\tau(b_i)=\false$ otherwise. If $\tau(b_i)=\true$, then mark all vertices in $U_i$, otherwise mark all vertices in $\overline{U}_i$.
In the end, $F[\tau]$ contains at least $\lceil(2^k+2)/2^k\rceil = 2$ variables from $U$, which form a cycle with $y$ and $z$ in $\inc(F[\tau])$.
This cycle is a contradiction to $B$ being a \sBDS of $F$.

\RuleKillingManyCycles*
As the previous rule is not applicable, every vertex $z \neq y$ is an interesting external killer for at most $2^{k-1}$ of these $\cC'$-cycles. Thus, no set of
interesting external killers of these $\cC'$-cycles of size at most $k$ excludes $y$. It follows that $y$ is in any \sBDS $B\subseteq \var'(F)$ of $F$ of size $k$.

\RuleTooManyCycles*
If none of Rules \ref{rule:SNoExtKiller}--\ref{rule:KillingManyCycles} applies, then $F$ has no \sBDS $B\subseteq \var'(F)$ of $F$ of size $k$. Indeed every vertex is an interesting external killer for at most $k\cdot 2^{k-1}$ $\cC'$-cycles,
but the number of $\cC'$-cycles is $\kextcycles = \kcycles-k = k^2 \cdot 2^{k-1} +1$.
\myqed \end{proof}

%% file: proofLemStrongcycles.tex
\begin{proof}
 The algorithm starts with $S^*=\emptyset$.
 For each choice $\mathfrak{C}$ among the $\binom{\kcycles}{\kcur}$ cycles to be killed externally, the algorithm executes one of the described rules.
 It computes a set $S$ such that every \sBDS of $F$ of size at most $\kcur$ respecting $\mathfrak{C}$ contains a variable from $S$.
 We set $S^*$ to be the union of all $S$ that are returned over all choices of cycles to be killed externally.
 As any \sBDS respects at least one such choice, $F$ has a \sBDS of size at most $\kcur$ containing at least one variable from $S^*$ if $F$ has a \sBDS of size at most $\kcur$.

 It remains to bound the size of $S^*$. The largest $S$ are returned by Rule \ref{rule:KillingManyCycles} and have size 2.
 As $\binom{\kcycles}{\kcur} \le (k^{2} 2^{k-1} + k+1)^k$, the lemma follows.
\myqed \end{proof}

%% file: proofThmStrong.tex
\begin{proof}
If $k\le 1$, our algorithm solves the problem exactly in polynomial time. Otherwise,
it invokes the algorithm from Theorem \ref{thm:Bodlaender} to either find a set of at least $\kcycles$ \vdcs, or
a \fvs of $G$ of size at most $\kfvs$.

In case it finds a \fvs of $G$ of size at most $\kfvs$, it uses Lemma \ref{lem:strongfvs} to compute a \sBDS of $F$ of size $k$ if one exists,
and it returns the answer.

In case it finds a set of at least $\kcycles$ \vdcs, it executes the procedure from Lemma \ref{lem:strongcycles}
to find a set $S^*$ of $O(k^{2k} 2^{k^2-k})$ variables such that any \sBDS of $F$ of size at most $k$ contains at least one variable from $S^*$.
The algorithm considers all possibilities that the Backdoor set contains every $x\in S^*$; there are $O(k^{2k} 2^{k^2-k})$ choices for $x$.
For each such choice, recurse on $F[x = \true]$ and $F[x = \false]$ with parameter $k-1$.
If, for any $x\in S^*$, both recursive calls return \sBDSs $B$ and $B'$, then return $B\cup B'\cup \set{x}$,
otherwise, return \textsc{No}. As $2^k-1 = 2\cdot (2^{k-1}-1)+1$, the solution size is upper bounded by $2^k-1$.
On the other hand, if at least one recursive call returns \textsc{No} for every $x\in S^*$, then $F$ has no \sBDS of size at most $k$.
\myqed \end{proof}